\documentstyle[prl,floats,twocolumn,aps]{revtex}

\begin{document}
\draft
\title{Quantum Communication with Correlated Nonclassical States}
\author{S. F. Pereira,* Z. Y. Ou,** and H. J. Kimble}
\address{Norman Bridge Laboratory of Physics 12-33 \\
California Institute of Technology\\
Pasadena, CA 91125}
\date{\today}
\maketitle

\begin{abstract}
Nonclassical correlations between the quadrature-phase amplitudes of two
spatially separated optical beams are exploited to realize a two-channel
quantum communication experiment with a high degree of immunity to
interception. For this scheme, either channel alone can have an arbitrarily
small signal-to-noise ratio (SNR) for transmission of a coherent
``message''. However, when the transmitted beams are combined properly upon
authorized detection, the encoded message can in principle be recovered with
the original SNR of the source. An experimental demonstration has achieved a
3.2 dB improvement in SNR over that possible with correlated classical
sources. Extensions of the protocol to improve its security against
eavesdropping are discussed.
\end{abstract}




\section{Introduction}

Principal motivations for the investigation of manifestly quantum or
nonclassical states of the electromagnetic field have been their possible
exploitation for optical communication$^{\cite{tak,yuen,cav}}$ and for
enhanced measurement sensitivity.$^{\cite{caves}}$ For example, relative to
a coherent state, the reduced quantum fluctuations associated with squeezed
and number states offer potential for improving channel capacity in the
transmission of information.$^{\cite{cav}}$ Squeezed states of light have
been widely employed to achieve measurement sensitivity beyond the standard
quantum limits in applications such as precision interferometry,$^{\cite
{xiao87}}$ the detection of directly encoded amplitude modulation,$^{\cite
{xiao88}}$ atomic spectroscopy,$^{\cite{pol}}$ and quantum noise reduction
in optical amplification.$^{\cite{ou93}}$ Likewise, nonclassical
correlations for the amplitudes of spatially separated beams have been
exploited in diverse situations, including demonstrations of the EPR paradox
for continuous variables,$^{\cite{ou92}}$ of quantum nondemolition detection
(QND),$^{\cite{roch,grangier98}}$ and of a quantum-optical tap.$^{\cite{poiz}%
}$

Within the broader setting of quantum information science (QIS), there has
been growing interest and important progress concerning the prospects for
quantum information processing with continuous quantum variables, including
universal quantum computation,$^{\cite{lloyd98a}}$ quantum error correction,$%
^{\cite{lloyd98b,sam98i,sam98ii}}$ and entanglement purification.$^{\cite
{plenio99,cirac99}}$ Theories for quantum teleportation of continuous
quantum variables in an infinite dimensional Hilbert space have been
developed,$^{\cite{vaidman,sam98a,ralph98,kurizki99}}$ including for broad
bandwidth teleportation$^{\cite{sam98b}}$ and for teleportation of atomic
wavepackets.$^{\cite{scott99a}}$ This formalism has also been applied to
super-dense quantum coding.$^{\cite{sam98c}}$ On an experimental front,
these developments in QIS led to the first {\it bona fide }demonstration of
quantum teleportation, which was carried out by exploiting nonclassical
states of light in conjunction with continuous quantum variables.$^{\cite
{furusawa98,fuchs99}}$

Against this backdrop the focus of attention in this article is optical
communication in two channels with quantum correlated light fields and the
associated quadrature amplitudes.$^{\cite{patent}}$ The goal is to explore
the extension of quantum cryptography from the usual setting of discrete
variables as pioneered by C. Bennett and colleagues$^{\cite{ben92a}}$ (e.g.,
photon polarization as in the experiments of Refs.\cite
{franson95,gisin00,hughes00,townsend98}) into the realm of continuous
quantum variables (e.g., the complex amplitude of the electromagnetic
field). Apart from our work, several related schemes for quantum
cryptography based upon continuous variables have recently been analyzed,
including a single-beam scheme with squeezed light$^{\cite{hillery99}}$ as
well dual-beam schemes with shared entanglement.$^{\cite{ralph99,reid99}}$
However, we stress at the outset that neither for our scheme nor for any of
these other protocols, can any claim about{\it \ absolute} security be made.
Rather, we suggest that these protocols (and suitable extensions thereof)
are worthy candidates for more detailed analyses. Such an undertaking would
involve various important matters of principle as well as practice for
continuous quantum variables, and might hopefully lead to security proofs
such as have recently emerged in the case of discrete variables.$^{\cite
{ben99,lo99,mayers99}}$

As illustrated in Figure 1, the basic idea in our scheme is to construct a
``transmitter'' which combines a coherent signal of amplitude $\epsilon /t$
(the ``message'') with the large fluctuating fields generated in
nondegenerate optical parametric amplification (the ``noise'').$^{\cite
{patent}}$ The ``message'' and the ``noise'' are superimposed at mirror $M$
with transmission coefficient $t$ $<<$ 1. Note that although each of the two
transmitted beams along channels ($A,B$) has large phase insensitive
fluctuations that are individually indistinguishable from a thermal source,$%
^{\cite{NOPA}}$ the quadrature-phase amplitudes of the two-beams can be
quantum copies of one another,$^{\cite{ou92,kim}}$ and in fact form an
entangled EPR state.$^{\cite{EPR,reid}}$ Hence proper subtraction of the
photocurrents at the ``receiver'' can result in the faithful reconstruction
of the encoded ``message'' even though the signal-to-noise ratios $R_{j}$ ($%
j $=$A,B$) during transmission are individually much less than one. Indeed,
in a lossless system with large parametric gain, the signal-to-noise ratio
of the reconstructed message $R_{t}$ can approach the signal-to-noise ratio $%
R_{0}$ of the original message ($\epsilon ^{2}/t^{2}$), which was written in
the transmitter as a coherent state before the mirror $M$ in Figure 1. Note
that the individual Channels $(A,B)$ have a high degree of immunity to
unauthorized interception since the signal-to-noise ratios $R_{A,B}$ in
these channels are each very small. Furthermore, any attempt to extract
information from the $(A,B)$ channels will reveal itself either by a
decrease in R$_{t}$ (classical extraction) or by an increase in the
fluctuations of the orthogonal quadrature amplitude (quantum extraction).

In addition to achieving a faithful reconstruction of the message
transmitted through M to the receiver, note that the scheme of Figure 1 also
preserves to a high degree the signal-to-noise ratio for the original
message beam that reflects from M. More specifically, for high gain and for
losses dominated by the transmission coefficient of M, the signal-to-noise
ratio R$_{r}$ for the reflected beam can approach R$_{0}$ for the original
message. In this limit, we then have that the information transfer
coefficient T $\equiv $ (R$_{r}$+R$_{t})$ / R$_{0}\rightarrow $ 2, where $%
0\leq $ T $\leq $ 1 for classical devices and 1 $<$ T $\leq $ 2 for
manifestly quantum or nonclassical situations.$^{\cite{poiz}}$ Hence the
scheme depicted in Fig.1 acts as a quantum optical tap in the fashion
originally discussed by Shapiro.$^{\cite{sha}}$ It provides a received (or
``tapped'') message with a signal-to-noise ratio equal to that of the input (%
$R_{t}/R_{0})\rightarrow 1$, while simultaneously transmitting an output
field with signal-to-noise ratio equal to that of the input $%
(R_{r}/R_{0})\rightarrow 1$.

Of course similar schemes for two-channel communication can be implemented
with correlated classical noise sources (i.e., thermal light), each with
large fluctuations which ``hide'' the message $\epsilon $ during
transmission. However, with classical sources of whatever type, only excess
fluctuations can be subtracted; the quantum fluctuations at the vacuum-state
level will remain unchanged and will enforce a noise floor for information
transmission and extraction. For the case illustrated in Figure 1, this
noise floor for the message at the receiver is given by the sum of
independent vacuum fluctuations from fields in channels ($A,B$) and sets a
fundamental noise level of ``2'' (with ``1'' as the individual vacuum-state
limits for the two channels). Here we adopt the usual convention for the
demarcation between classical and nonclassical correlations in terms of the
behavior the Glauber-Sudarshan phase-space function.$^{\cite{kim}}$ Hence
for the case illustrated in Figure 1 but with classical input fields, the
signal-to-noise ratio $R_{t}^{\prime }$ for the detected message at the
receiver is given by $R_{t}^{\prime }\simeq |\epsilon |^{2}\ll R_{t}$. In
fact, for classical inputs, we have that $T^{\prime }\equiv R_{t}^{\prime
}+R_{r}^{\prime }\leq 1$, and the system no longer functions as a quantum
optical tap. Furthermore, the individual channels ($A,B$) are not protected
from unauthorized eavesdropping, since information can be extracted from
these channels with impunity for classical noise much greater than the
vacuum-state limit.

Apart from these considerations related to secure communication and quantum
optical tapping, the configuration of Figure 1 can also be viewed as a means
to realize super-dense quantum coding$^{\cite{densecoding}}$ for continuous
quantum variables.$^{\cite{sam98c}}$ Here, the message $\epsilon /t$ is
again encoded at the mirror $M$, but now in a single channel corresponding
to one component of the entangled EPR state (e.g., channel $A$). This
combination of the message and the fluctuations from one component of the
NOPA are transmitted to the receiving station where they are combined with
the second component of the entangled output of the NOPA that has been
independently transmitted (e.g., along channel $B$). The signal is then
decoded by combining the outputs of the two channels in a fashion similar to
that shown in Figure 1 as discussed in more detail in Ref.\cite{sam98c}. The
principal distinctions between this dense coding scheme and the
aforementioned dual channel arrangement are (1) the message is encoded in a
single component of the entangled EPR beam instead of symmetrically in both
and (2) the received beams from paths $(A,B)$ must be physically recombined,
with the phases of the local oscillators $(A,B)$ at the receiving station
offset by $\frac{\pi }{2}$. Recall that for dense coding in its canonical
form,$^{\cite{densecoding}}$ no signal modulation is applied to the second
(i.e., channel $B$) component of the entangled state, so that it carries no
information by itself.

In subsequent sections of this paper, we describe in more detail the
implementation of this general discussion about quantum communication with
correlated nonclassical fields. In our experiment, we have been able to
demonstrate an improvement in signal-to-noise ratio by a factor of 2.1 over
that possible with any classical source (that is, 10 $log[R_{t}/R_{t}^{%
\prime }]=3.2$ dB) and have succeeded in suppressing the noise of the
difference photocurrent $i_{-}\equiv i_{A}-i_{B}$ below that associated with
the vacuum fluctuations of even a single beam, thus making possible
transmission with $|\epsilon |^{2}$ $<1$. Quantum dense coding would thereby
be enabled with the aforementioned changes in the overall experimental
protocol. We conclude with a discussion of possible extensions for enhanced
security against unauthorized eavesdropping.

\section{Implementation by Nondegenerate Parametric Amplification}

As illustrated in Figure 1, correlated nonclassical states for our work are
generated by a nondegenerate optical parametric amplifier (NOPA) that
produces orthogonally polarized but frequency degenerate signal and idler
beams for channels ($A,B$). We emphasize that these beams represent a
realization of the entangled state originally discussed by Einstein,
Podolsky, and Rosen.$^{\cite{EPR,reid}}$ For the original EPR state, there
exist perfect correlations both in position and momentum for two massive
particles. In the optical case, the quadrature amplitudes of the
electromagnetic field play the roles of position and momentum with a finite
degree of correlation for finite NOPA gain, as has been experimentally
demonstrated$^{\cite{ou92}}$ and exploited to realize quantum teleportation.$%
^{\cite{furusawa98}}$

A coherent-state ``message'' of total amplitude $\epsilon /t$ is encoded in
equal measure onto these entangled EPR beams by orienting its polarization
at 45$^{\circ }$ with respect to the signal and idler polarizations at the
mirror $M$ of Figure 1. To obtain a quantitative statement of the
performance of this system, we must include the finite gain of the amplifier
as well as various passive losses, which together limit the degree of
correlation that can be exploited for communication. Following the analysis
of Ref.\cite{ou92}, we find that the SNR $R_{j}(\Omega )$ for the individual
signal and idler photocurrents for propagation and detection in the presence
of overall channel efficiency $\xi $ is given by $R_{j}(\Omega )=\xi {{%
\epsilon ^{2}}/{2G_{q}(\Omega )}}$, where $G_{q}(\Omega )$ is the detected
quantum-noise gain of the amplifier which can be determined experimentally
from measurements of the spectral densities $\Psi _{A,B}(\Omega )$ for the
fluctuations of photocurrents for signal and idler beams alone at either
detector. Relative to the frequency of the optical carrier determined by the
down-conversion process in the NOPA, the frequency $\Omega $ specifies the
Fourier components of the quadrature-phase amplitudes of signal and idler
fields as well as of the coherent field $\epsilon $.$^{\cite{kim}}$ Note
that $\xi $ ($0\leq \xi \leq 1$) incorporates the cavity escape efficiency
for our NOPA, the propagation efficiency from the NOPA to the detectors, and
the homodyne and quantum efficiencies of the balanced detectors themselves.$%
^{\cite{ou92}}$

Although the individual fluctuations for channels ($A,B$) give rise to a
level $G_{q}(\Omega )>1$, (that is, greater than the vacuum-state limit of
either beam alone), these large fluctuations are correlated in a
nonclassical manner and hence can be eliminated by proper choice of the
quadrature amplitudes detected at ($A,B$). As shown in Ref.\cite{ou92},
there is a continuous set of such amplitudes with minimum variance for their
difference requiring only that the quadrature-phase angles ($\theta
_{A},\theta _{B})$ satisfy $\theta _{A}+\theta _{B}=2p\pi $ ($p$ = integer).
Denoting one such pair by ($X_{A},X_{B}$), we have that 
\[
\langle (X_{A}(\Omega )-X_{B}(\Omega ))(X_{A}(\Omega ^{\prime
})-X_{B}(\Omega ^{\prime }))\rangle 
\]
\begin{equation}
=V_{-}(\Omega )\delta (\Omega +\Omega ^{\prime }),  \label{variance}
\end{equation}
where $V_{-}(\Omega )$ is a variance which quantifies the degree of
correlation between $(X_{A},X_{B}$). Explicit expressions for both $%
V_{-}(\Omega )$ and $G_{q}(\Omega )$ are given in Ref.\cite{ou92}. For
propagation and detection in the presence of loss, we introduce the
quantities ($V_{-}^{d},G_{q}^{d}$) which refer to the variance and quantum
noise gain for fictitious fields having propagated with total loss $(1-\xi )$%
, where the spectral density of the photocurrent fluctuations $\Phi
_{-}(\Omega )$ is proportional to $V_{-}^{d}(\Omega )$. Hence, the SNR $%
R_{d} $ for detection of the message via $i_{-}$ is given by $R_{d}={{2\eta
\epsilon ^{2}}/{V_{-}^{d}(\Omega )}}$, where $\eta $ accounts for the
propagation and detection efficiency for the message from the mirror M to
the photocurrent $i_{A,B}$. Without discussing the general case, here we
note simply that for efficient propagation and detection with $(1-\xi )<<1$
and for near threshold operation with (analysis frequency $\Omega )<<$
(cavity linewidth $\Gamma )$, then $G_{q}^{d}(\Omega )\rightarrow 1+{\frac{1%
}{2}}(\Gamma /\Omega )^{2}\xi $, while $V_{-}^{d}(\Omega )\rightarrow
2(1-\xi )<<1$, so that $R_{d}(\Omega )\rightarrow \eta \epsilon ^{2}/(1-\xi
) $. Hence in the ideal case with $\eta \rightarrow $1 and $\xi \rightarrow
(1-{|t|}^{2})$, with t as the amplitude transmission coefficient of mirror
M, we find that the reconstructed message is recovered with the same SNR
with which it was originally encoded (namely $R_{d}\rightarrow \epsilon
^{2}/t^{2} $), while the fluctuations in the individual channels become
arbitrarily large ($R_{A,B}\sim \Omega ^{2}\epsilon ^{2}/\Gamma
^{2}\rightarrow 0$ for $\Omega /\Gamma \rightarrow 0$).

As for the performance as an optical tap, note that the transfer coefficient
associated with the {\em detected} message at the receiver and with the
reflected output field is given by $T_d = (R_d+R_r) /R_0$, where $R_d$ is
related to $R_t$ by way of the propagation and detection efficiency $\eta$
from M to the photocurrents at the receivers. In the present case, we have
that 
\begin{equation}
T_d= {\frac{{\vert r \vert ^2} }{{U_-^r (\Omega)}}} + {\frac{{\ 2 \eta \vert
t \vert ^2} }{{V_-^d(\Omega)}}},
\end{equation}
with $\vert r \vert ^2$ as the reflectivity of mirror M $({\vert r \vert}%
^{2} + {\vert t \vert}^{2}$=1) and $U_-^r(\Omega)$ as the variance of the
reflected field. Hence in the ideal case with $\xi \rightarrow (1-\vert t
\vert^2)$, with V$_-^d(\Omega) \rightarrow 2(1-\xi)$ and with $U_-^r(\eta) =
\vert r \vert ^2$, we have that $R_d \rightarrow R_t$ and $T_d \rightarrow 2$%
. Thus in addition to providing large quantum fluctuations for secure
transmission, the system also acts as a quantum optical tap with a nearly
ideal transfer coefficient $T$.

In fact the system can be considered as a realization of the scheme for
quantum tapping that was originally suggested by Shapiro.$^{\cite{sha}}$ To
see this more clearly, recall that the projection of signal and idler fields
along the $45^{\circ}$ polarization direction of the message beam results in
a squeezed field.$^{\cite{kim}}$ Hence, from the perspective of Ref.$\cite
{sha}$, we are ``tapping'' the original message field by injecting squeezed
light into the normally open (or vacuum) port of mirror M. The use of the
output of a nondegenerate parametric amplifier allows us subsequently to
decompose this squeezed plus coherent field into individually noisy signal
and idler fields at polarizer P for transmission.

\section{Experimental Setup and Results}

The general scheme for our experimental implementation of these ideas is
shown in Figure 1, where frequency degenerate but orthogonally polarized
signal and idler beams are generated by Type II down-conversion in a
subthreshold optical parametric oscillator formed by a folded cavity
containing an $a$-cut crystal of potassium titanyl phosphate (KTP) that
provides noncritical phase matching at 1.08$\mu $m. The crystal is 10mm
long, is anti-reflection coated for both 1.08$\mu $m and 0.54$\mu $m, and
has a measured harmonic conversion efficiency of $6\times 10^{-4}$/W
(single-pass) for this geometry. The total intracavity passive losses at 1.08%
$\mu $m are 0.3\% and the transmission coefficient of mirror M1 is 3\%. The
amplifier is pumped by green light at 0.54 $\mu $m generated by external
frequency doubling of a frequency-stabilized, TEM$_{00}$-mode Nd:YAP laser.$%
^{\cite{ouol}}$ The subthreshold oscillator acts as a narrow-band amplifier
(NOPA) which is locked to the original laser frequency with a weak
counter-propagating beam. Simultaneous resonance for the orthogonally
polarized signal and idler fields is achieved by adjusting the temperature
of the KTP crystal around 60$^{\circ }$C with milliKelvin precision. The
pump field at 0.54 $\mu $m is itself resonant in a separate and
independently locked build-up cavity (enhancement $\sim $ 5x).

As we have demonstrated in our previous experiments, $^{\cite{ou92}}$ the
orthogonally polarized signal and idler fields generated by the NOPA
individually are fields of zero mean values and exhibit large phase
insensitive fluctuations. It is in the midst of this noise that we now hide
a ``message'', with this coherent field being combined with the signal and
idler fields at the highly reflecting mirror $M$ shown in Figure 1 ($%
(1-t^{2})\simeq 0.99$). The coherent beam is injected at 45$^{\circ }$ with
respect to signal and idler polarizations and is frequency shifted by $%
\Omega _{0}/2\pi =1.1$MHz (single-side band) from the primary laser
frequency with the help of a pair of acoustooptic modulators, which are
gated ``on'' and ``off'' to provide information encoded for transmission.
The noisy but correlated signal and idler beams together with the coherent
information are then separated by a polarizer $P$, transmitted independently
over the two channels ($A,B$), and then directed to two separate balanced
homodyne detectors for measurements of their individual quadrature-phase
amplitudes and their mutual correlations. The local oscillators for the two
balanced homodyne detectors originate from the laser at 1.08$\mu $m; their
phases can be independently controlled by mirrors mounted on piezoelectric
transducers. The spectral densities of the photocurrents for the two
channels ($A=$ signal, $B=$ idler) are defined by 
\begin{equation}
\Psi _{A,B}(\Omega )=\int {<i_{A,B}(t)i_{A,B}(t+\tau )>e^{i\Omega \tau }d{%
\tau }}
\end{equation}
and are recorded by a RF spectral analyzer, as is the spectral density 
\begin{equation}
\Phi _{-}(\Omega )=\int {<i_{-}(t)i_{-}(t+\tau )>e^{i\Omega \tau }d{\tau }}
\label{phiminus}
\end{equation}
for the difference photocurrent $i_{-}\equiv i_{A}-i_{B}$.

In Figures 2 and 3 we present results from a series of measurements of these
various spectral densities. First of all, in Figure 2a, trace {\it i} gives
the spectral density $\Psi _{A}$ for channel A alone with an injected
``message'' and with the amplifier turned on to generate large $(\sim $ 7
dB) phase insensitive noise above the vacuum-state level $\Psi _{0A}$
(indicated by a dashed line in Figure 2) for the signal beam. A similar
trace is obtained for the spectral density $\Psi _{B}$. By contrast, trace 
{\it ii} in Figure 2a gives the spectral density $\Phi _{-}$ for the
difference photocurrent $i_{-}$, with the phases of the local oscillators
adjusted for minimum noise and maximum coherent signal. In this trace, the
coherent message that was completely obscured in trace {\it i} emerges with
high signal-to-noise ratio. Note that in trace {\it ii} the correlated
quantum fluctuations for signal and idler fields are subtracted to
approximately 0.4 dB below the vacuum-noise level $\Psi _{0A}$ of the signal
beam alone (and likewise for the idler), indicating an improvement in SNR
over a conventional single-channel communication scheme with a classical
light source.

To complete the discussion, we present in Figure 2b results obtained with
the amplifier turned off (that is, uncorrelated vacuum-state inputs for
signal and idler fields which are combined with the coherent ``message''
information at mirror $M$). Trace {\it i} shows the result for the signal
beam alone ($\Psi _{A}$), where again the noise floor $\Psi _{0A}$ is from
the vacuum fluctuations of the signal beam; a similar trace is obtained for
the idler beam $(\Psi _{B})$. Trace {\it ii} gives the corresponding result
for $\Phi _{-}$ for the combined signal and idler photocurrents when the
amplifier is off. Note that this trace represents the best possible SNR with
which the encoded information can be recovered when correlated classical
noise sources are employed since here the (uncorrelated) vacuum fluctuations
of signal and idler beams set an ultimate noise floor 3 dB above $\Psi _{0A}$
(that is, $\Phi _{0-}=2\Psi _{0A}$).$^{\cite{inf}}$ On comparing traces {\it %
ii} in Figures 2a and 2b, we see that the correlated quantum fluctuations of
signal and idler fields brought about by parametric amplification result in
an improvement in SNR of 3.2 dB relative to that possible with classical
noise sources.

The improvement in SNR with correlated quantum fields over classical fields
in our two-channel communication scheme can be of utility especially when
the message is so weak that the SNR is poor for transmission with correlated
classical sources (that is, for the case where vacuum noise dominates the
encoded message). This situation is illustrated in Figure 3, where we plot $%
\Phi _{-}$ for the two cases without (trace {\it i}) and with (trace {\it ii}%
) correlated quantum fields.$^{\cite{inf}}$ Relative to Figure 2, here the
coherent beam has been attenuated resulting in a smaller SNR for the
``message''. Indeed in trace {\it i}, this information is ``buried'' by the
vacuum noise $\Phi _{0-}$ associated with independent vacuum fluctuations in
channels $A$ and $B$; recovery of the encoded information is poor. On the
other hand, as shown in trace {\it ii}, when correlated quantum fields are
employed, there is a reduction in the noise floor by more than 3 dB which
makes possible improved recovery of the encoded information, with the
recovery here limited by losses in propagation and detection.$^{\cite{ou92}}$

As for the actual performance with respect to optical tapping, our system
falls far short of the projected possibilities discussed in the preceding
section because of an unfortunate mismatch between the transmissivity $%
|t|^{2}$ for mirror M and the overall system efficiency $\xi $. In
quantitative terms, recall that the transfer coefficient $T$ for encoding
information from the input beam to the reflected and transmitted beams at M
is given by $T\equiv (R_{r}+R_{t})/R_{0}$ whereas the transfer coefficient
for the detected message photocurrent and the reflected signal field is $%
T_{d}\equiv (R_{r}+R_{d})/R_{0}$ as given explicitly in Eq. (2). For the
propagation and detection efficiencies in our experiment $(\xi \simeq $ 0.65
and $\eta \simeq $ 0.75), these transfer coefficients are optimized for
mirror transmission ${|t|}^{2}\sim $ 0.5 for M. In our arrangement we have
instead $|t|^{2}=0.01$, with the inferred result that $T_{d}\simeq 1.02$,
which is only marginally in the quantum domain.

In the experiment described here, the receiver uses a local oscillator (LO)
that originates from the fundamental frequency of the same laser that
generated the pump beam for the NOPA. This LO is necessary for proper
detection of the quadrature amplitudes of the nonclassical beams and of the
message, since it provides a phase reference that follows phase fluctuations
of the NOPA's pump beam. In practice, as the stability of the available
lasers improve, one should consider schemes for which the measurement is
carried out with nominally independent lasers for the LO and for the source.
For example, one might employ a stabilized laser diode as a reference to
phase lock lasers both at the sender and at the receiver, where the laser
diode could be widely distributed through optical fibers. Alternatively,
Ralph has analyzed a scheme in which the local oscillators are transmitted
and recovered as part of the overall protocol.$^{\cite{ralph99}}$

\section{Comparison with Other Dual Beam Schemes}

It is perhaps obvious that the degree of immunity to interception for a two
channel scheme such as we have discussed is related to the degree of excess
fluctuations for each individual beam. For the demonstration in Ref.\cite
{man}, the excess noise used to ``hide'' the encoded information in each
beam comes from some artificial unrelated source. Unfortunately such
uncorrelated excess fluctuations also add noise to the coincidence signal in
the recovery of the ``message,'' even though the added noise scales
differently as a function of photon number for single-beam measurements
(linearly) and for dual beam measurements (quadratically). Hence larger
background noise which better ``hides'' the encoded information also brings
larger added noise in the extraction of the ``message.'' Because of the
quadratic dependence on the total photon number for the extra noise added in
coincidence detection, this scheme is best suited to low light level
transmission, as demonstrated in the pioneering experiment by Hong et al.$^{
\cite{man}}$

The situation is quite different for the quadrature-phase amplitudes of the
correlated signal and idler fields generated by the NOPA. As the NOPA is
pumped harder and the threshold for parametric oscillation is approached,
the gain of the amplifier increases, as do the excess fluctuations of the
signal and idler fields. However, the correlation between the fluctuations
of the signal and idler beams also improves, giving rise to even better SNR
for the recovered signal. The key point is that the large fluctuations in
the signal and idler beams needed for immunity to interception are intrinsic
and do not add extra noise to the recovered signal but, on the contrary,
serve to reduce the noise in $i_{-}$ as the gain of the amplifier increases.
In the end, the SNR for the recovered message is arbitrated by the imperfect
correlation resulting from finite gain and from passive losses in
propagation and detection. On the other hand, this dependence provides a
powerful means to detect eavesdropping because unauthorized extraction of
signal or idler fields from channels $A$ or $B$ results in a reduction of
the detected correlation and hence an increase in the noise floor of the
recovered message. Note that unauthorized extraction of information from
both channels by way of a quantum optical tap$^{\cite{sha}}$ or a quantum
nondemolition measurement$^{\cite{grangier98}}$ can likewise be detected
because of the unavoidable increase of fluctuations for the orthogonal
quadrature-phase amplitudes ($\theta _{A,B}+\pi /2$) of the two channels.
Furthermore, these quantum eavesdropping schemes can be defeated in large
measure by random switching of the phases of the message, signal, and idler
beams as discussed below.

Our system also offers advantages with respect to the (classical) digital
Vernan cipher, where a message is decomposed in two correlated random
signals and transmitted over two one-way channels. Although this system
seems to be similar to ours in the sense that is also secure provided the
eavesdropper has access to one channel only, the situation is different if
the eavesdropper can split a small fraction of both channels since in the
classical case, this can be done without the knowledge of the receiver.
However, in our system the eavesdropper cannot choose arbitrarily the
reflectivity of any ``beamsplitter'' used for extraction from the two
channels since in the quantum case, the fraction of the beams extracted
should be big enough so that the signal-to-noise ratio for the intercepted
message is greater than one. But if this is the case, then unavoidable extra
``noise'' added to the transmitted beams by the open port of the
``beamsplitter'' degrades the signal-to-noise ratio of the message at the
legitimate receiver, thus revealing the unauthorized intervention during
transmission.

One might attempt to circumvent this difficulty by employing a quantum
extraction procedure, such as quantum nondemolition detection$^{\cite
{grangier98}}$ of the quadrature amplitudes in Channels $(A,B)$. Although
the signal-to-noise ratio $R_{d}$ at the receiver would not in this case be
degraded by an ideal eavesdropper, the unauthorized intervention could
nonetheless be discovered because of the injection of large fluctuations
(``back-action'' noise) in the quadrature orthogonal to that in which signal
information is stored, as previously noted.

\section{Extensions via Random Phase Switching}

One way an eavesdropper {\it Eve}\ could access the signal and idler beams
without the knowledge of the legitimate receiver is if she can intercept
both channels completely, detect in the same manner as does the legitimate
receiver (i.e., {\it Eve} should also have access to a local oscillator
phase stable with respect to that of sender and receiver) and retransmit the
beams in the same way as the legitimate sender. Because of this possibility,
our protocol as described is certainly not secure, in contrast to the
protocols for discrete variables.$^{\cite{ben99,lo99,mayers99}}$ However, we
suggest that simple extensions of our protocol might lead to significant
enhancements in security.

If the goal were to achieve quantum key distribution, one idea is to make
straightforward adaptations of the protocols introduced by Bennett and
colleagues for the discrete case, as in Ref.\cite{hillery99,ralph99,reid99}.
Here, we propose that the sending station ({\it Alice}) and receiving
station {\it (Bob}) make random choices for the set of phases of the
coherent message beam, as well as for the signal and idler beams. Recall
that the variance $V_{-}(\Omega )$ of Eq.\ref{variance} is the minimum
possible and applies only for the choice of quadrature-phase angles $(\theta
_{A},\theta _{B})$ for the signal and idler beams that satisfy $\theta
_{A}+\theta _{B}=2p\pi $ ($p$ = integer). For definiteness, assume the
following two choices.

\begin{enumerate}
\item  $(\theta _{A}^{0},\theta _{B}^{0})$, with $\theta _{A}^{0}+\theta
_{B}^{0}=0$ and corresponding quadrature amplitudes $(X_{A},X_{B})$.

\item  $(\theta _{A}^{\pi /2}=\theta _{A}^{0}+\frac{\pi }{2},\theta
_{B}^{\pi /2}=\theta _{B}^{0}+\frac{\pi }{2})$ and corresponding quadrature
amplitudes $(Y_{A},Y_{B})$.
\end{enumerate}

In the first case, the minimum variance $V_{-}(\Omega )$ results for the
combination $(X_{A}-X_{B})$, while in the second case, the combination $%
(Y_{A}+Y_{B})$ has minimum variance. This is because $Y_{B}\rightarrow
-Y_{B} $ is equivalent to the shift $\theta _{B}^{\pi /2}\rightarrow \theta
_{B}^{\pi /2}+\pi $, so that $\theta _{A}^{\pi /2}+\theta _{B}^{\pi /2}+\pi
=2\pi $.

With these definitions, {\it Alice} at the sending station (randomly) makes
one of two choices.

\begin{enumerate}
\item  {\it Phase} $0$ -- Set the quadrature-phase angles $(\theta
_{A},\theta _{B})$ to $(\theta _{A}^{0},\theta _{B}^{0})$ and the phases $%
\beta _{A,B}=\beta _{A,B}^{0}$ for the coherent message beam $|\alpha
\rangle =|\alpha |\exp [i\beta ]$ corresponding to the $X$ quadratures of $%
(A,B)$.

\item  {\it Phase} $\frac{\pi }{2}$ -- Set $(\theta _{A},\theta _{B})$ to $%
(\theta _{A}^{\pi /2},\theta _{B}^{\pi /2})$ and $\beta _{A,B}=\beta
_{A,B}^{\pi /2}=\beta _{A,B}^{0}\pm \frac{\pi }{2}$ corresponding to the $Y$
quadratures.
\end{enumerate}

The encoded message (which could consist of $|\alpha |=[a_{0},a_{1}]$ for a
binary transmission) is sent to {\it Bob's} receiving station precisely as
in Figure 1. {\it Bob} must then choose the appropriate phases $(\phi
_{A},\phi _{B})$ for his local oscillators $(LO_{A},LO_{B})$ to detect
quadrature amplitudes such that the spectral density $\Phi _{-}(\Omega )$
for the difference photocurrent $i_{-}\equiv i_{A}-i_{B}$ is minimized and
the signal maximized. In the case {\it Phase} $0$, denote the local
oscillator settings as $(\phi _{A}^{0},\phi _{B}^{0})$, in correspondence to
the detection of $(X_{A},X_{B})$ with minimum variance $V_{-}(\Omega )$. On
the other hand, for the case {\it Phase} $\frac{\pi }{2}$, the local
oscillator phases $(\phi _{A},\phi _{B})\rightarrow (\phi _{A}^{\pi /2},\phi
_{B}^{\pi /2})=(\phi _{A}^{0}+\frac{\pi }{2},\phi _{B}^{0}+\frac{3\pi }{2})$%
, in correspondence to the detection of $(Y_{A},-Y_{B})$ with minimum
variance. In both cases, the encoded message would be recovered with maximum
signal-to-noise ratio. Note that precisely such a switching protocol was
implemented in our prior experiment of Ref.\cite{ou92} with results as
stated for the variances.

Of course, {\it Bob }does not know in advance which choice $[0,\frac{\pi }{2}%
]$ {\it Alice} will have made for any given transmission. Hence, he makes a
random selection between the alternatives $(\phi _{A}^{0},\phi _{B}^{0})$
and $(\phi _{A}^{\pi /2},\phi _{B}^{\pi /2})$, recovering the message in
some cases but not others. After a series of transmissions, {\it Alice} and 
{\it Bob} communicate publicly about their choice of bases, keeping
measurement results only when their choices coincide.

Now, if an eavesdropper {\it Eve} attempts to intervene (either by a
strategy of partial tapping or by one of complete interception and
re-broadcast), she will necessarily increase the noise level and error rate
at {\it Bob's }receiving station. The random switching of the phases $%
(\theta _{A},\theta _{B})$ by {\it Alice} forces {\it Eve} to make a guess
as to the correct quadratures $(\delta _{A},\delta _{B})$ to be detected.
Having made a choice, information about the orthogonal quadrature is lost.
Of course, rather than homodyne detection, she could choose to employ
heterodyne detection to gain information about the full complex amplitude.
However, relative to homodyne detection, heterodyne detection brings a
well-known penalty of a $3$dB reduction in signal-to-noise ratio.$^{\cite{AK}%
}$

While it is beyond the scope of the current paper to make any claims about
the quantitative limits to the information that {\it Eve} might access or
about the absolute ability of {\it Alice }and {\it Bob }to detect her
presence, we do suggest that these would be interesting questions to
investigate. There are certainly intervention strategies beyond those that
we have mentioned that a cunning {\it Eve} would want to consider, such as
an adaptive strategy for adjusting the phases $(\delta _{A},\delta _{B})$
during the duration of the transmission of any given message.$^{\cite
{wiseman95}}$ Likewise, in any real-world setting, overcoming the
deleterious effects of losses in propagation from {\it Alice} to {\it Bob}
will be a overriding consideration. The question of preserving the
entanglement of the initial EPR\ state in the face of such losses is a
fascinating one for continuous quantum variables. Although initial attempts
have been made to develop error correcting quantum codes for continuous
variables,$^{\cite{lloyd98b,sam98i,sam98ii}}$ no adequate solution seems to
yet have been found. Finally, it would be of interest to analyze the case
where only one of the two correlated beams is sent to {\it Bob}, with then 
{\it Alice }retaining the other.

\acknowledgments

We gratefully acknowledge the comments of J.H. Shapiro who pointed out the
connection of our experiment to Ref.\cite{sha}, of S. L. Braunstein and H.
Mabuchi for critical discussions, and of one of the referees who brought to
our attention the Vernon cipher. This work was supported by the Office of
Naval Research, by the National Science Foundation, and by DARPA via the
QUIC administered by the Army Research Office.

\begin{figure}[tbp]
\caption{Principal components of the experiment showing the ``transmitter'',
where a message $\protect\epsilon $ is combined with noise fields from a
nondegenerate optical parametric amplifier (NOPA) at mirror $M$. The
orthogonally polarized signal and idler beams are separated by polarizer $P$%
, and then propagate along independent channels ($A,B$) to two separate
balanced homodyne detectors that form the ``receiver''. In the figure,
arrows represent coherent amplitudes of various fields, while the shaded
circles are meant to indicate their fluctuations.}
\label{exp}
\end{figure}

\begin{figure}[tbp]
\caption{(a) Signal recovery with correlated quantum states in channels ($%
A,B $). Trace i gives the spectral density $\Psi_A(\Omega)$ of photocurrent
fluctuations for channel $A$ alone as a function of time. Trace ii is the
spectral density $\Phi_-(\Omega)$ for the combined photocurrent $i_-=i_A-i_B$
from both channels; here the ``message'' (coherent beam chopped on and off)
clearly emerges. (b) Signal recovery with uncorrelated vacuum fluctuations.
Again, trace i gives $\Psi_A(\Omega)$ (channel $A$ only) while trace ii
gives $\Phi_-(\Omega)$ (combined photocurrent $i_-$). In (a) and (b) the
vacuum-state level $\Psi_{0A}$ for channel $A$ (signal beam only) and $%
\Phi_{0-}$ for the combined photocurrent $i_-$ (dual beam) are shown as
dashed lines. Note that $\Psi_{0A}$ lies 15 dB above the electronic noise
floor. Spectrum analyzer acquisition parameters are as follows: resolution
bandwidth = 100 kHz, video bandwidth = 100 Hz, analysis frequency $\Omega_0
/2\protect\pi = 1.1$ MHz, and sweep time = 300 ms. }
\label{big}
\end{figure}

\begin{figure}[tbp]
\caption{Spectral density of the photocurrent fluctuations $\Phi_-(\Omega)$
for $i_-=i_A-i_B$ for the case when the on-off modulation of the message is
encoded with small SNR. Trace i $-$ Uncorrelated vacuum fluctuations. Trace
ii $-$ Correlated quantum fluctuations in channels ($A,B$). The vacuum-state
limits $\Psi_{0A}$ and $\Phi_{0-}$ are indicated. Acquisition parameters are
as in Fig. 2. }
\label{small}
\end{figure}


\begin{references}
\bibitem[*]{byline1}  Present address: Delft University of Technology,
Faculty of Applied Sciences - Optics Research Group, Lorentzweg 1, 2628 CJ
Delft, The Netherlands.

\bibitem[**]{byline2}  Present address: Department of Physics, Indiana
University-Purdue University at Indianapolis, 402 N Blackford St.,
Indianapolis, IN 46202.

\bibitem{tak}  H. Takahasi, Adv. Commun. Syst.{\bf 1}, 227 (1965).

\bibitem{yuen}  H. P. Yuen and J. H. Shapiro, IEEE Trans. Inform. Th. {\bf %
IT-24}, 657 (1978).

\bibitem{cav}  C. M. Caves and P. D. Drummond, Rev. Mod. Phys. (1994); H. P.
Yuen and M. Ozawa, Phys. Rev. Lett. {\bf 70}, 363 (1993).

\bibitem{caves}  C. M. Caves, Phys. Rev. {\bf D23}, 1963(1981) and Phys.
Rev. {\bf D26}, 1817(1982); C. M. Caves et al., Rev. Mod. Phys. {\bf 52},
341(1980).

\bibitem{xiao87}  M. Xiao, L. A. Wu, and H. J. Kimble, Phys. Rev. Lett. {\bf %
59}, 278 (1987); P. Grangier, R. E. Slusher, B. Yurke, and LaPorta, Phys.
Rev. Lett. {\bf 59}, 2153 (1987).

\bibitem{xiao88}  M. Xiao, L.A. Wu, and H. J. Kimble, Opt. Lett. {\bf 13},
476 (1988).

\bibitem{pol}  E.S. Polzik, J. Carri, and H. J. Kimble, Phys. Rev. Lett. 
{\bf 68}, 3020 (1992).

\bibitem{ou93}  Z. Y. Ou, S. F. Pereira, and H. J. Kimble, Phys. Rev. Lett. 
{\bf 70}, 3239 (1993).

\bibitem{ou92}  Z. Y. Ou, S. F. Pereira, H. J. Kimble, and K. C. Peng, Phys.
Rev. Lett. {\bf 68}, 3663 (1992); Z. Y. Ou, S. F. Pereira, and H. J. Kimble,
Appl. Phys. {\bf B55}, 265 (1992).

\bibitem{roch}  J. F. Roch, G. Roger, P. Grangier, J. M. Courty, and S.
Reynauld, Appl. Phys. {\bf B55}, 291 (1992).

\bibitem{grangier98}  For a review, see P. Grangier, J. A. Levenson, and
J.-P. Poizat, Nature {\bf 396}, 537 (1998).

\bibitem{poiz}  J. Ph. Poizat and P. Grangier, Phys. Rev. Lett. {\bf 70},
271 (1993).

\bibitem{lloyd98a}  S. Lloyd and S. L. Braunstein, quant-ph/9810082.

\bibitem{lloyd98b}  S. Lloyd and J. J. E. Slotine, Phys. Rev. Lett. {\bf 80}%
, 4088 (1998).

\bibitem{sam98i}  S. L. Braunstein, Phys. Rev. Lett. {\bf 80}, 4084 (1998).

\bibitem{sam98ii}  S. L. Braunstein, Nature {\bf 394}, 47 (1998).

\bibitem{plenio99}  S. Parker, S. Bose, and M. B. Plenio, quant-ph/9906098.

\bibitem{cirac99}  L.-M. Duan, G. Giedke, J. I. Cirac, and P. Zoller,
quant-ph/9912017.

\bibitem{vaidman}  L. Vaidman, Phys. Rev. {\bf A49}, 1473 (1994).

\bibitem{sam98a}  S. L. Braunstein and H. J. Kimble, Phys. Rev. Lett. {\bf 80%
}, 869 (1998).

\bibitem{ralph98}  T. C. Ralph and P. K. Lam, Phys. Rev. Lett. {\bf 81},
5668 (1998).

\bibitem{kurizki99}  T. Opatrny, G. Kurizki, and D.-G. Welsch,
quant-ph/9907048.

\bibitem{sam98b}  P. van Loock, S. L. Braunstein, and H. J. Kimble, Phys.
Rev. {\bf A} (accepted, 2000); quant-ph/9902030.

\bibitem{scott99a}  A. S. Parkins and H. J. Kimble, Journal of Optics B -
Quantum and Semiclassical Optics {\bf 1}, 496 (1999), available as
quant-ph/9904062; and submitted, 1999, quant-ph/9909021.

\bibitem{sam98c}  S. L. Braunstein and H. J. Kimble, Phys. Rev. {\bf A}
(accepted, 2000); quant-ph/9810082.

\bibitem{furusawa98}  A. Furusawa, J. Sorensen, S. L. Braunstein, C. Fuchs,
H. J. Kimble, and E. S. Polzik, Science {\bf 282}, 706 (1998).

\bibitem{fuchs99}  S.\ L.\ Braunstein, C.\ A.\ Fuchs, H.\ J.\ Kimble,
Journal of Modern Optics {\bf 47}, 267 (2000), available as quant-ph/9910030.

\bibitem{patent}  United States Patent $5,339,182$, ``Method and Apparatus
for Quantum Communication Employing Nonclassical Correlations of
Quadrature-Phase Amplitudes,'' issued August 16, 1994, H. J. Kimble, Z. Y.
Ou, and S. F. Pereira.

\bibitem{ben92a}  C. H. Bennett, G. Brassard, and A. K. Ekert, Sci. Am. {\bf %
267}, 50 (1992).

\bibitem{franson95}  J. D. Franson, B. C. Jacobs, Electronics Letters {\bf 31%
}, 232 (1995).

\bibitem{gisin00}  G. Ribordy, J. D. Gautier, N. Gisin, O. Guinnard, H.
Zbinden, J. Mod. Optics {\bf 47}, 517 (2000).

\bibitem{hughes00}  R. J. Hughes, G. L. Morgan GL, and C. G. Peterson, J.
Mod. Optics {\bf 47}, 533 (2000).

\bibitem{townsend98}  P. D. Townsend, Opt. Fiber Technology {\bf 4}, 345
(1998).

\bibitem{hillery99}  M. Hillery, quant-ph/9909006.

\bibitem{ralph99}  T. C. Ralph, quant-ph/9907073.

\bibitem{reid99}  M. D. Reid, quant-ph/9909030.

\bibitem{ben99}  For a recent review of security proofs in quantum
cryptography, see C. H. Bennett and P. W. Shor, Science {\bf 284}, 747
(1999).

\bibitem{lo99}  H.-K. Lo and H. F. Chau, Science {\bf 283}, 2050 (1999).

\bibitem{mayers99}  D. Mayers, quant-ph/9802025.

\bibitem{NOPA}  S. M. Barnett and P. Knight, J. Opt. Soc. Am. {\bf B2}, 467
(1985).

\bibitem{kim}  H. J. Kimble, in {\it Fundamental Systems in Quantum Optics},
ed. by J. Dalibard, J. M. Raimond, and J. Zinn Justin (Elsevier, Amsterdam,
1992). 545ff.

\bibitem{EPR}  A.\ Einstein, B.\ Podolsky, N.\ Rosen, Phys.\ Rev.\ {\bf 47},
777 (1935).

\bibitem{reid}  M.\ D.\ Reid and P.\ D.\ Drummond, Phys.\ Rev.\ Lett.\ {\bf %
60}, 2731 (1988); M.\ D.\ Reid, Phys.\ Rev.\ A {\bf 40}, 913 (1989).

\bibitem{sha}  J. H. Shapiro, Opt. Lett. {\bf 5}, 351 (1980).

\bibitem{densecoding}  C.\ H.\ Bennett and S.\ J.\ Wiesner, Phys.\ Rev.\
Lett.\ {\bf 69}, 2881 (1992).

\bibitem{ouol}  Z. Y. Ou, S. F. Pereira, E. S. Polzik, and H. J. Kimble,
Opt. Lett. {\bf 17}, 640 (1992).

\bibitem{inf}  In Figs.(2,3), $\Phi _{-}$=1.9 $\Psi _{0,A}$ (2.8 dB above)
due to a slight imbalance (0.1 dB) between the ($A,B$) detectors and to a
small contribution (0.1 dB) from detector thermal noise.

\bibitem{man}  L. Mandel, J. Opt. Soc. Am. {\bf B1},108 (1984); C. K. Hong,
S. R. Friberg, and L. Mandel, Appl. Opt. {\bf 24}, 3877 (1985).

\bibitem{AK}  E.\ Arthurs and J.\ L.\ Kelly Jr., Bell. Syst. Tech. J. April,
725 (1965).

\bibitem{wiseman95}  H. M. Wiseman, Phys.\ Rev.\ Lett.\ {\bf 75}, 4587
(1995); H. M. Wiseman and R. B. Killip, Phys.\ Rev.\ {\bf A57}, 2169
(1998).\newpage
\end{references}
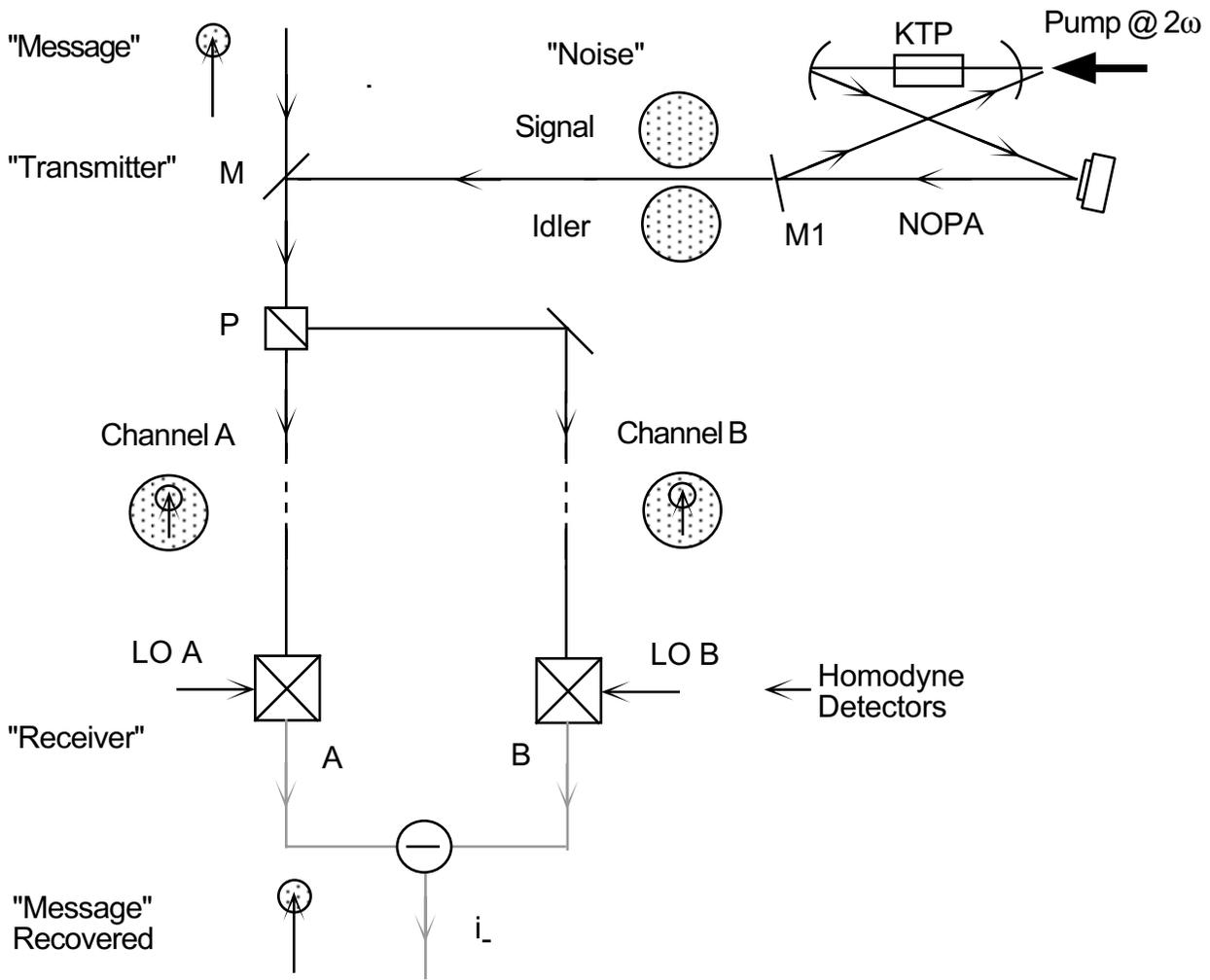
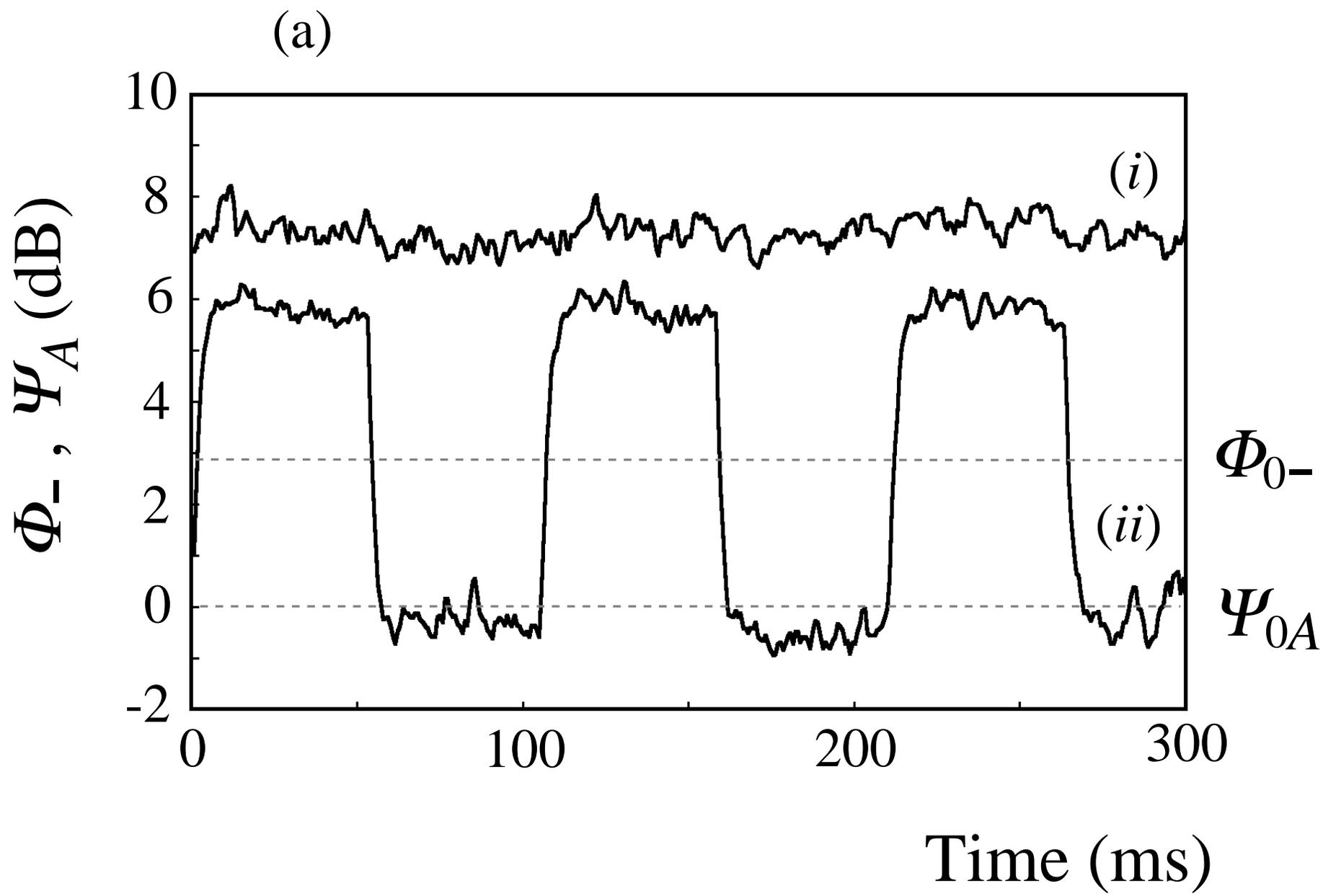
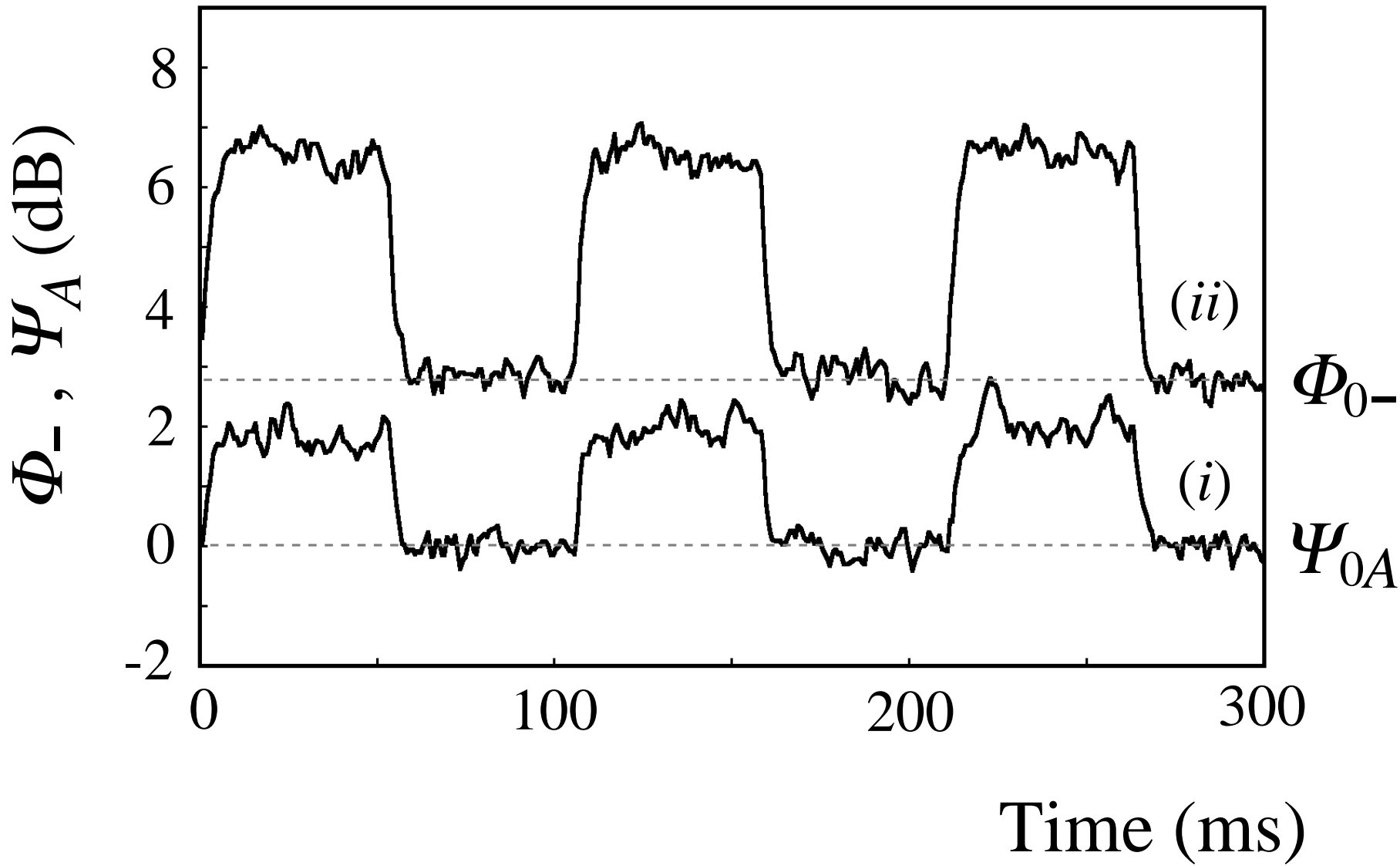
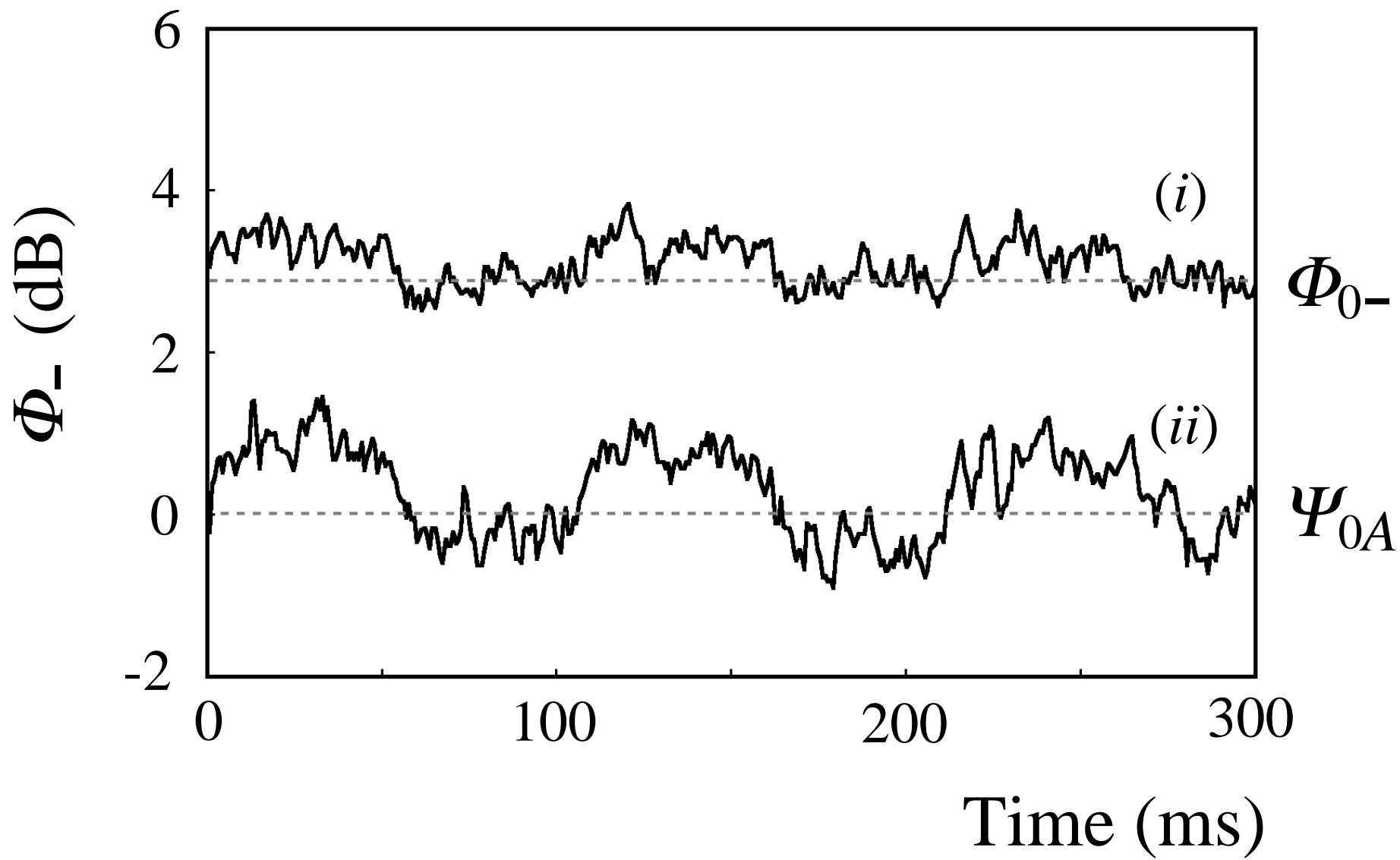
\end{document}